\documentclass[12pt,twoside]{article}
\usepackage{amsmath}
\newcommand{\rd}[1]{\mathop{\mathrm{d}#1}}

\newcommand{\fract}[2]{{\textstyle\frac{#1}{#2}}}
\newcommand{\grad}{\vec\nabla}

\newcommand{\nA}{non-Abelian}
\newcommand{\CS}{Chern-Simons}
\newcommand{\CSt}{Chern-Simons term}
\newcommand{\Cpr}{Clebsch pa\-ra\-me\-ter\-iza\-tion}

\newcommand{\pr}{para\-me\-ter\-iza\-tion}
\newcommand{\prd}{para\-me\-ter\-ized}

\newcommand{\mn}{{\mu\nu}}
\newcommand{\pp}[1]{\partial_{#1}}

\newcommand{\numeq}[2]{\begin{equation}
#2
\label{#1}
\end{equation}}
\newcommand{\refeq}[1]{(\ref{#1})}

\let\vec\boldsymbol
\let\eps\varepsilon
\let\epsilon\varepsilon
\let\phi\varphi

\begin{document}
 
\title{John Bell's Observations on the Chiral Anomaly\\
and Some Properties of Its Descendants}
\author{R. Jackiw\\
\small\it Center for Theoretical Physics\\ 
\small\it Massachusetts Institute of Technology\\ 
\small\it Cambridge, MA 02139-4307}
\date{\small MIT-CTP\#3043 \quad Typeset in \LaTeX\ by M. Stock\\[1ex]
\normalsize\itshape Bell Memorial Meeting, Vienna, Austria, November 2000}
\maketitle

\abstract{\noindent
John Bell's emphasis of the essential ambiguities in anomaly calculations is recalled. 
Some descendants of the anomaly are reviewed.
\null}

\pagestyle{myheadings}
\markboth{\small {\it R. Jackiw}}{\small  John Bell's Observations on the Chiral
Anomaly}
\thispagestyle{empty}

\section{John Bell and the Chiral Anomaly}

I expect that most everyone in this audience knows that John Bell codiscovered the
mechanism of anomalous symmetry breaking in quantum field theory. Indeed, our
paper on this subject~\cite{ref1} is his (and my) most-cited work. The symmetry
breaking in question is a quantum phenomenon that violates the correspondence
principle; it arises from the necessary infinities of quantum field theory. Over the
years it has become evident that theoretical/mathematical physicists are not the only
ones to acknowledge this effect. Nature makes fundamental use of the anomaly in at
least two ways: the neutral pion's decay into two photons is controlled by the
anomaly~\cite{ref1,ref2} and elementary fermions (quarks and leptons) arrange
themselves in patterns such that the anomaly cancels in those channels to which
gauge bosons -- photon, W,  Z  -- couple~\cite{ref3}. (There are also phenomenological
applications of the anomaly to collective, as opposed to fundamental, physics -- for
example, to edge states in the quantum Hall  effect.) Beyond physics, in mathematics
one finds closely related structures, such as Atiyah-Singer index theory, zero modes
of Dirac operators, Chern-Pontryagin characteristics of gauge fields, \CS\ secondary
characteristics. The mathematical ideas were developed at nearly the same time as
the physical ones, and this unexpected conjunction between physics and mathematics
seeded joint activity that flourishes to this day. 

Once it was appreciated that the anomaly is not merely an obscure pathology of
quantum field theory, but reflects an as-yet-to-be-understood wrinkle in the
field-theoretic description of Nature, many people wrote many papers providing
various and alternative derivations of the result. But not John Bell. It seems that all
he had to say on the subject was contained in our first joint paper. He did  follow
the subsequent developments and elaborations, and he commented on them to me
whenever we met in Geneva or Cambridge. 

In his characteristically diffident manner, he did not always support the various
fancy elaborations, and he remained skeptical about their value. On the contrary, he
insisted on one statement, which already appeared in our original paper, but which
perhaps had not been forcefully enunciated, so I want to call attention to it here. 

Our original analysis concerns the correlation function for three currents -- one axial
vector and two vector currents -- which is given in lowest order by the fermionic
triangle graph. With massless fermions, this correlation function should be
transverse in all three channels, the axial vector and both vectors, as a consequence
of various symmetries in the theory. The anomaly  manifests itself  in that any
evaluation of the relevant diagram is ambiguous up to a local term, owing to the
underlying infinities of the quantum field theory. Moreover, no matter how one fixes
the ambiguity, the transversality conditions fail -- the calculated amplitude is not
transverse in all three channels. Thus the full extent of symmetry, which would
ensure full transversality, is broken. While specific choices for resolving the
ambiguity allow transversality in some (but not all three) channels, John Bell always
insisted that there is no intrinsic way to select a ``correct'' result. And this opinion
informed his criticism of alternative derivations of the anomaly, which usually
preserve vector transversality at the expense of axial vector transversality.
Whenever we discussed yet another new approach to the anomaly, he always wanted
to verify that there remained the possibility of obtaining a variety of results. If this
freedom were not present, he would dismiss the rederivation as too restrictive. 

One alternative viewpoint did appeal to John Bell. The lack of transversality in a
particular channel can be ascribed to an anomalous nonconservation of
the appropriate ``symmetry'' current, with the nonvanishing anomalous current
divergence determined by the gauge fields with which the fermions interact. But
from another perspective, one could state that  transversality fails because the
commutators between the relevant current operators differ from the corresponding
Poisson brackets of their classical antecedents --  the commutators contain additional,
quantal (anomalous) contributions. (Indeed, this is the point of view one must adopt
in the interaction picture, where operator equations do not see the interaction.) The
form for the anomalous commutators is uniquely determined by the triangle graph,
yet the transversality conditions retain their ambiguity, because they continue to
reflect the arbitrariness in the local contribution to the graph. This gives an appealing
algebraic characterization of the anomaly~\cite{ref4}. Either point of view signals
absence of the expected symmetries and exposes a breakdown of the
correspondence principle.

John Bell's insistence that the triangle graph supports a variety of anomalies is not
merely a pedantic nicety. In fact it has a physical realization in Gerard 't~Hooft's
calculation of fermion-number nonconservation in the standard model~\cite{ref5},
where the vectorial fermion number current in the triangle graph carries the
anomaly, and the chiral current is anomaly free~\cite{ref6}. 
\vspace*{-\bigskipamount}

\section{Descendants of the Anomaly}

The axial anomaly, that is, the departure of transversality of the 3-fermion current
correlation function, involves
${}^*\!FF$, an expression constructed from the gauge fields to which the fermions
couple. Specifically, in the Abelian case one encounters
\numeq{eq1}{
{}^*\!F^\mn  F_\mn = \fract12 \eps^{\mn\alpha\beta} F_\mn F_{\alpha\beta} =
-4
\vec E\cdot \vec B
 }
where $F_\mn$ is the covariant electromagnetic tensor 
\begin{subequations}
\numeq{eq2a}{
F_\mn = \pp\mu A_\nu - \pp \mu A_\nu
 }
while $\vec E$ and $\vec B$ are the electric and magnetic fields
\numeq{2a}{
 E^i = F^{io}\ ,\quad
 B^i = -\fract12 \eps^{ijk} F_{jk}\ .
}
\end{subequations}
The non-Abelian generalization reads
\numeq{eq3}{
{}^*\! F^{\mn a} F_\mn^a   = \fract12 \eps^{\mn\alpha\beta} F_\mn^a
F_{\alpha\beta}^a 
}
where $F_\mn^a$ is Yang-Mills gauge field strength 
\numeq{eq4}{
F_\mn^a = \pp\mu A_\nu^a - \pp \nu A_\mu^a + f^{abc} A_\mu^b A_\nu^c
}
and $a$ labels the components of the gauge group, whose structure constants
are~$f^{abc}$. 

The quantity $ {}^*\!FF$ is topologically interesting. Its integral over 4-space is
quantized, and measures the topological class (labeled by integers) to which the
vector potential~$A$ belongs. Consequently, the 4-volume integral of $ {}^*\!FF$ is a
topological invariant and we expect that, as befits a topological invariant, it should be
possible to present $ {}^*\!FF$ as a total derivative, so that its 4-volume integral
becomes converted by Gauss' law into a surface integral, sensitive only to long
distance, global properties of the gauge fields. 
That the total derivative form for ${}^*\!FF$ holds is seen when $F_\mn$ is expressed
in terms of potentials. In the Abelian case, we use \refeq{eq2a} and find immediately
\numeq{eq5}{
\fract12  {}^*\!F^\mn F_\mn  = \pp \mu \bigl(\eps^{\mu\alpha\beta\gamma}
A_\alpha \pp\beta A_\gamma \bigr)\ .
 }
For the non-Abelian fields, \refeq{eq4} establishes the desired result:
\numeq{eq6}{
\fract12 {}^*\!F^{\mn a} F_\mn^a   = \pp\mu \eps^{\mu\alpha\beta\gamma}
\bigl( A_\alpha^a\pp\beta A_\gamma^a + \fract13 f^{abc} A_\alpha^a A_\beta^b
A_\gamma^c
\bigr)\ .
}
The quantities whose divergence gives ${}^*\!FF$ are called \CS\ terms. By
suppressing one dimension they become naturally defined on a 3-dimensional
manifold (they are 3-forms), and we are thus led to consider the \CS\ terms in their
own right~\cite{ref7}:
\begin{align}
\mathrm{CS}(A) &= \eps^{ijk} A_i \pp j A_k & &\text{(Abelian)}\label{eq7}\\
\mathrm{CS}(A) &= \eps^{ijk} \bigl(A_i^a \pp j A_k^a +\fract13 f^{abc} A_i^a A_j^b
A_k^c\bigr)
 & &\text{(non-Abelian).}\label{eq8}
\end{align}

The 3-dimensional integral of these quantities is again topologically interesting. When
the
\nA\
\CSt\ is evaluated on a pure gauge, \nA\ vector potential
\numeq{eq9}{
A_i = g^{-1} \pp i g
} 
the 3-dimensional volume integral of $\mathrm{CS}(g^{-1} \partial  g)$ measures the
topological class (labeled by integers) to which $g$ belongs. The integral in the
Abelian case -- the case of electrodynamics -- is called the magnetic helicity $\int
\rd{^3 r} \vec A\cdot
\vec B$,
$\vec B = \grad
\times \vec A$, and measures the linkage of magnetic flux lines. An analogous
quantity arises in fluid mechanics, with the local fluid velocity $\vec v$ replacing
$\vec A$, and the vorticity $\vec\omega = \grad\times\vec v$ replacing~$\vec B$. Then
the integral $\int \rd{^3 r} \vec v\cdot \vec\omega$ is called kinetic
helicity~\cite{ref8}.

I shall not review here the many uses to which the \CSt s, Abelian and \nA,
introduced in~\cite{ref7}, have been put. The applications range from the
mathematical characterization of knots to the physical description of electrons in the
quantum Hall effect~\cite{ref9},  vivid evidence for the deep significance of the \CS\
structure and of its antecedent, the chiral anomaly.

Instead I pose the following question: Can one write the \CSt\ as a total derivative, so
that (as befits a topological quantity) the spatial volume integral becomes a surface
integral. An argument that this should be possible is the following: The
\CSt\ is a 3-form on 3-space, hence it is maximal and its exterior derivative vanishes
because there are no 4-forms on 3-space. This establishes that on 3-space  the \CSt\
is closed, so one can expect that it is also exact, at least locally, that is, it can be
written as a total derivative. 
Of course, such a representation for the \CSt\ requires expressing the
potentials in terms of ``pre-potentials'', since the 
 formulas
\refeq{eq7}, \refeq{eq8} show no evidence of derivative structure.
[Recall that the total derivative formulas \refeq{eq5}, \refeq{eq6} for the axial
anomaly also require using potentials to express~$F$.]

There is a physical, practical reason for wanting the Abelian \CSt\  to be a total
derivative. It is known in fluid mechanics that there exists an obstruction to
constructing a Lagrangian for Euler's fluid equations, and this obstruction is just the
kinetic helicity  \hbox{$\int \rd{^3 r} \vec v\cdot \vec\omega$}, that is, the volume
integral of the Abelian \CSt, constructed from the velocity 3-vector~$\vec v$. This
obstruction is removed when the integrand is a total derivative, because then the
kinetic helicity volume integral is converted to a surface integral by Gauss' theorem.
When the integral obtains contributions only from a surface,  the obstruction
disappears from the 3-volume, where the fluid equation acts~\cite{ref10}. 

It is easy to show that the  Abelian \CSt\ can be presented as a total derivative. We
use the \Cpr\ for  a 3-vector~\cite{ref11}
\numeq{eq10}{
\vec A = \grad\theta + \alpha\grad\beta\ .
}
This nineteenth-century \pr\ of the a 3-vector $\vec A$ in terms of the prepotentials
($\theta$, $\alpha$, $\beta$) is an alternative to the usual transverse/longitudinal
\pr. In modern language it is a statement of Darboux's
theorem that the 1-form $A_i \rd{r^i}$ can be written as $\rd \theta + \alpha
\rd\beta$~\cite{ref12}. With this
\pr\ for $\vec A$, one sees that the Abelian \CSt\ is indeed a total derivative:
\begin{align}
\mathrm{CS}(A) &= \eps^{ijk} A_i \pp j A_k\label{eq11}\\
&=  \eps^{ijk} \pp i \theta\pp j \alpha \pp k \beta\nonumber\\
&= \pp i \bigl( \eps^{ijk} \theta\pp j \alpha \pp k \beta\bigr)\ .\nonumber
\end{align}

When the \Cpr\ is employed for $\vec v$ in the fluid dynamical context, the
situation is analogous to the force law in electrodynamics. While the Lorentz equation
is written in terms of field strengths, a Lagrangian formulation needs potentials from
which the field strengths are reconstructed. Similarly, Euler's equation involves the
velocity vector~$\vec v$, but in a Lagrangian for this equation the velocity must be
parameterized in terms of the  prepotentials $\theta$, $\alpha$, and~$\beta$. 

In a natural generalization of the above, one asks whether a \nA\ vector potential
can also be \prd\ in such a way that the \nA\ \CSt~\eqref{eq8} becomes a total
derivative. We have answered this question affirmatively and we have found
appropriate prepotentials that do the job~\cite{ref10,ref13,ref14}, but the details of
the construction are too technical to be presented here. We hope that our \nA\
generalization of the \Cpr\ will be as interesting and useful as the Abelian one.
\vspace*{-\bigskipamount}

\end{document}